\documentclass{iau}
\usepackage{ulem}
\usepackage{amsmath}
\usepackage{graphicx}
\usepackage{caption}
\usepackage{subfigure} 
\usepackage{multirow}
\usepackage{natbib}
\usepackage[font=footnotesize,labelfont=bf]{subcaption}
\begin{document}

\lefttitle{Colombo, et. al.}
\righttitle{Activity Cycles}

\jnlPage{1}{7}
\jnlDoiYr{2025}
\doival{10.1017/xxxxx}

\aopheadtitle{Proceedings IAU Symposium}
\editors{C. Sterken,  J. Hearnshaw \&  D. Valls-Gabaud, eds.}

\title{Activity Cycles in Solar-Type Stars}

\author{
P. D. Colombo$^{1}$,
C.A. Peralta$^{2,3}$,
A. P. Buccino$^{1,2}$,
C. G. Oviedo$^{1}$,
R. Ibañez Bustos$^{4}$,
C. F. Martinez$^{5}$,
M. Flores Trivigno$^{6}$, 
F. Minotti$^{2,3}$
\&
P. Mauas$^{1,2}$
}
\affiliation{$^1$Instituto de Astronomía y Física del Espacio, CONICET--UBA, Argentina; 
$^2$Departamento de Física, Facultad de Ciencias Exactas y Naturales, UBA, Argentina;
$^3$Instituto de Física Interdisciplinaria y Aplicada, UBA--CONICET, Argentina;
$^4$Observatoire de la Côte d’Azur, Francia;
$^5$Observatorio Astronómico de Córdoba, UNC, Argentina;
$^6$Instituto de Ciencias Astronómicas, de la Tierra y del Espacio, CONICET, Argentina}

\begin{abstract}

Systematic observational studies, mainly focused on solar-type stars in the F to early K range, have detected magnetic activity cycles analogous to the solar one.  When analyzing the relationship between the length of these cycles and stellar rotation rate, a non-monotonic dependence is observed. Activity periods are discretely distributed into two distinct branches: inactive and active. Intriguingly, some stars with rotation rates faster than the Sun exhibit two coexisting activity cycles, with periods simultaneously falling on both the inactive and active branches. In this work, we analyze the time series of the Mount Wilson index of a set of 44 stars, covering a range of rotation periods (sub-solar, solar, and super-solar), to identify single or double activity cycles. We detected activity cycles in 19 of the analyzed stars, with 7 of these exhibiting double-cycle behavior. In a complementary approach, we also analyze the activity cycles simulated for identical solar twins with rotation periods both shorter and longer than the Sun.

\end{abstract}
\maketitle

\section{Introduction}

The magnetic fields  responsible for the “stellar activity”  are generated by the interaction between convection and stellar rotation (stellar dynamo). Therefore, the rotation period is a critical parameter in the characterization of the stellar chromospheric activity. Moreover, empirical relationships have shown that the rotation rate of solar-type stars is directly correlated with the length of the activity cycle. In particular, this relationship does not seem to be univocal, and the periods of activity are distributed in two branches (inactive and active), and several solar-type stars even present two coexisting activity cycles distributed in two branches simultaneously. This double cycle could be the result of the local action of plasma turbulence ($\alpha$ effect) operating simultaneously on the surface and in the tachocline of the star. Interestingly, the Sun does not lie on either of these branches, occupying an intermediate and apparently atypical position in such diagrams. This peculiar location of the solar cycle could indicate that the solar dynamo is currently in transition. Along these lines, it has been noted that stars with rotation periods between 23 and 30 days are rarely reported as cyclic.

In order to provide observational evidence to this discussion, in this work we present a preliminary study of the long-term activity of a set of 44 solar-type stars, with rotation periods shorter, similar, and longer than the Sun. To do this, we employ our own spectra obtained at the Complejo Astronómico el Leoncito (CASLEO)\textbf{\footnote{Based on data acquired at Complejo Astronómico El Leoncito, operated under agreement between the Consejo Nacional de Investigaciones Científicas y Técnicas de la República Argentina and the National Universities of La Plata, Córdoba and San Juan.}} under the observing programme HK$\alpha$ Project and other public databases.
To complement this observational approach, we are able to build non-linear axisymmetric dynamo models, which can reproduce multiple activity cycles in solar-type stars. In the present work, we present the dynamo simulations of a hypothetical solar twin with rotation periods shorter and longer than the Sun.

\section{Sample Selection}

This work focuses on solar-type stars originally observed by the HK$\alpha$ Project, a sample characterized by a broad range of rotation periods, from 7.78 to 71.00 days (as detailed in Table 2 of \cite{Cincunegui07}).

With the aim of detecting long cycles (active branch) and short cycles (inactive branch) in solar-type stars with a range of rotation periods between 3 and 60 days (subsolar, solar, and supersolar), in this work we study a sample of stars with rotation periods within this range that were repeatedly observed throughout the HK$\alpha$ Project. Therefore, taking into account stellar parameters, cadence, and time span of observations, we restricted the sample to 44 stars. Table \ref{resumen} lists the sample of stars studied that presented activity cycles (19 stars), their main parameters, the observational data from CASLEO, and the results obtained in this work. The rotation periods of the stars in this list are taken exclusively from the literature \cite{Cincunegui07,Metcalfe10,Metcalfe13,Egeland15}.

\section{Long-Term Activity}

To compile the long-term activity records of our sample stars, we employed CASLEO spectra, public ESO data from HARPS and FEROS spectrographs\footnote{\url{https://archive.eso.org/cms/data-portal.html}}, and Mount Wilson indexes reported in \citep{Baum22}. The \textit{Mount Wilson index} $S$ \citep{Vaughan78} was calculated following \cite{Ibanez20}, with HARPS and FEROS indices calibrated to $S$ using \cite{Lovis11} and \cite{Jenkins08}, respectively. CASLEO indices were intercalibrated with quasi-simultaneous $S$ measurements from HARPS (preferably) or FEROS, and in some cases with values from \cite{Baum22}.

Long-term magnetic activity cycles were searched using the Generalized Lomb-Scargle (GLS) periodogram \citep{Zechmeister09}, consistent with the approach used in previous studies \citep{Ibanez19a,Ibanez19b,Ibanez20}. A star was classified as cyclic if its periodic variability reached a significance threshold of FAP$<$10\%. Following \cite{Colombo24}, all other stars were designated as irregular variables or constants.

As an example of cyclic stars with double activity cycles, we present the K3V star HD 32147 with a rotation period of 47.4 days \citep{Cincunegui07}, and an effective temperature of 4550 $\mathrm{K}$ \citep{Baum22}.

\begin{figure}[hbt!]
\centering
\includegraphics[width=0.6\columnwidth]{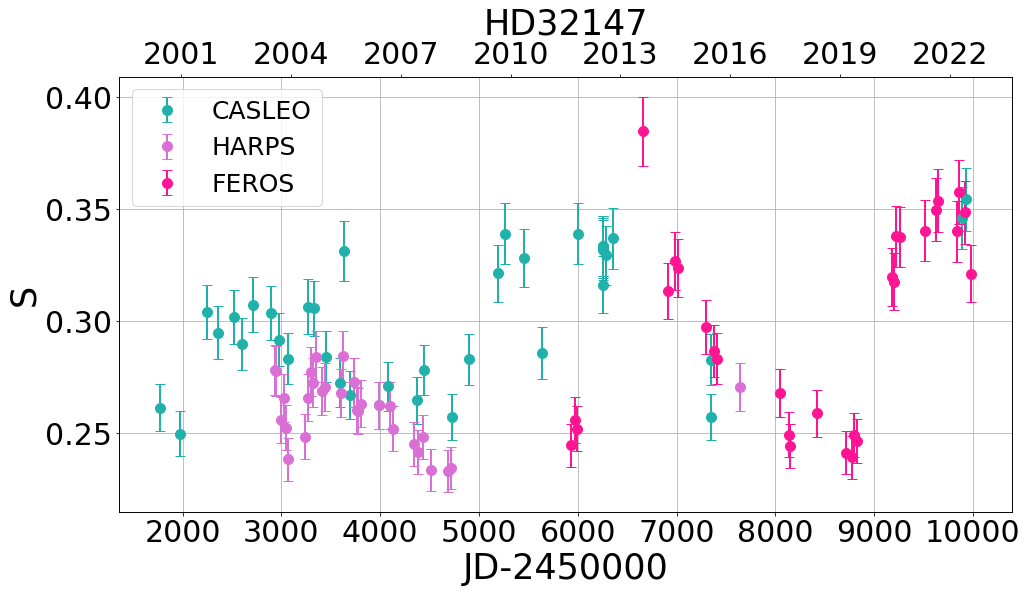}
\caption{$S$-index of HD 32147 between 2000--2023.}
\label{hd32147ST}
\end{figure}

Figure \ref{hd32147ST} shows the $S$-index time series built with 38 CASLEO spectra obtained between 2000 and 2022, and 264 HARPS and FEROS spectra taken between 2004 and 2023, both monthly  averaged. 

\begin{figure}[hbt!]
  \centering
  \makebox[0pt][c]{%
    \scalebox{1.15}{
      \begin{minipage}{0.48\textwidth}
        \centering
        \includegraphics[width=\linewidth]{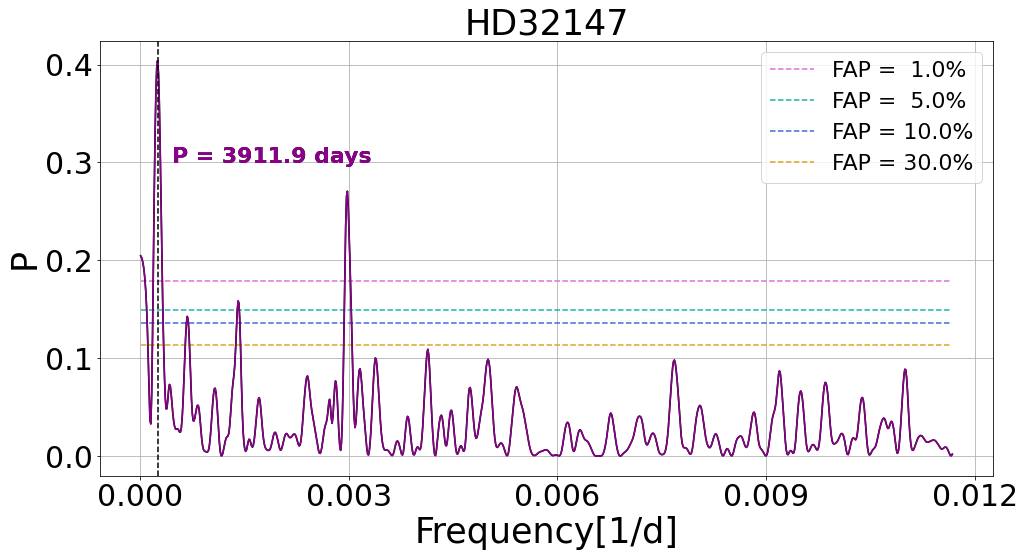}
       \captionof{subfigure}{First activity cycle}
        \label{hd32147P}
      \end{minipage}\hspace{0.02\textwidth}%
      \begin{minipage}{0.48\textwidth}
        \centering
        \includegraphics[width=\linewidth]{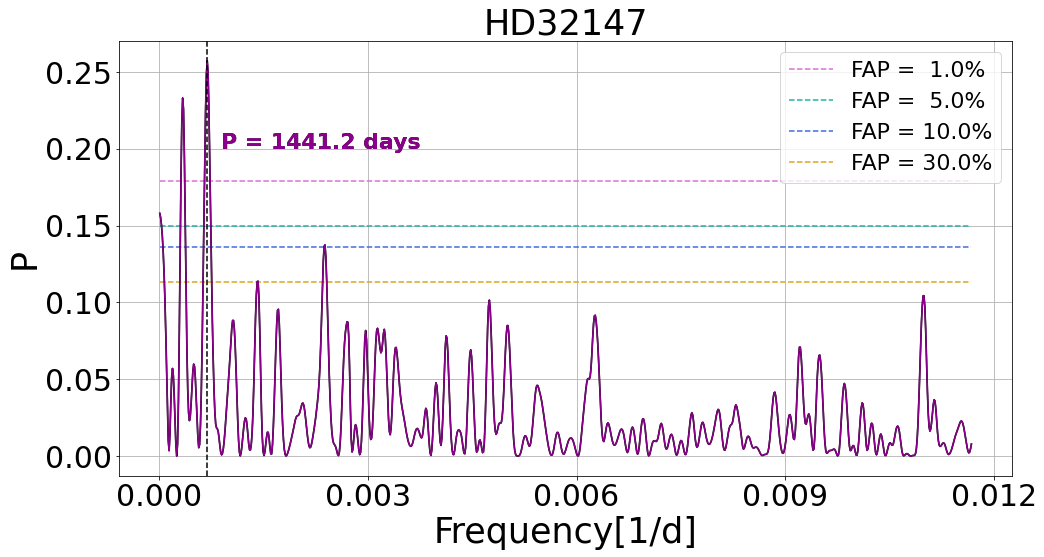}
       \captionof{subfigure}{Second activity cycle}
        \label{hd32147P2}
      \end{minipage}%
    }
  }
  \caption{GLS periodograms of HD 32147 whith the detections of the first activity cycle of (3911.9 $\pm$ 215.5) days and the second activity cycle of (1441.23 $\pm$ 33.99) days.}
  \label{P}
\end{figure}

We computed the GLS periodogram in Figure \ref{P}a for the $S$ index in Figure \ref{hd32147ST}, which reveals a significant peak at $P$ = (3911.9 $\pm$ 215.5) days $\sim$ (10.7 $\pm$ 0.6) years, with FAP = 3.4$\times$10$^{-9}$.

After removing from the original time series the harmonic component of this primary period, we computed the periodogram in Figure \ref{P}b still shows a secondary cycle at $P$ = (1441.23 $\pm$ 33.99) days $\sim$ (3.95 $\pm$ 0.095) years, with FAP = 9.4$\times$10$^{-5}$.

\section{Simulated Activity Cycles in Identical Solar Twins with Varying Rotation Rates}

In order to clarify the atypical position of the Sun in the $P_{\mathrm{cyc}}$--$P_{\mathrm{rot}}$ diagrams \citep{BohmVitense07,Metcalfe16}, we applied a large-scale axisymmetric dynamo model with feedback effects \citep{Sraibman19} to six solar twins. The formulation of this model, along with an example of its application to the solar dynamo, is fully described in \cite{Sraibman19}. This model, which allows us to obtain equations for the mean values of the fields and velocities, is derived by applying the formalism developed in \cite{Minnoti00} to the equations of magnetohydrodynamics.

\begin{figure}[hbt!]
  \centering
\makebox[0pt][c]{%
  \scalebox{1.1}{%
    \begin{minipage}{0.48\textwidth}
      \centering
      \includegraphics[width=\linewidth]{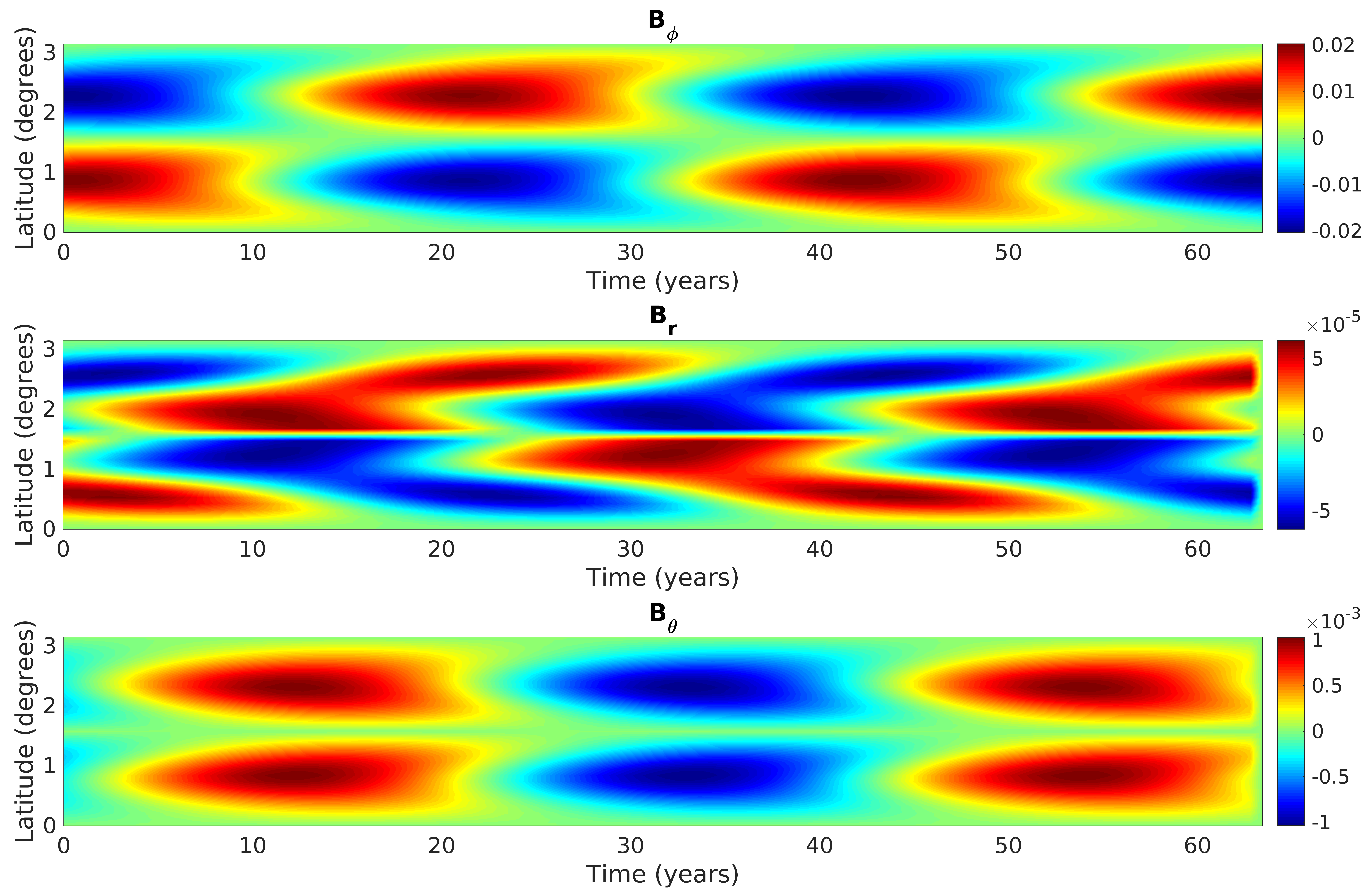}
      \captionof{subfigure}{Case $0.5\,\omega_\odot$}
      \label{fig:campo05w}
    \end{minipage}\hspace{0.02\textwidth}%
    \begin{minipage}{0.48\textwidth}
      \centering
      \includegraphics[width=\linewidth]{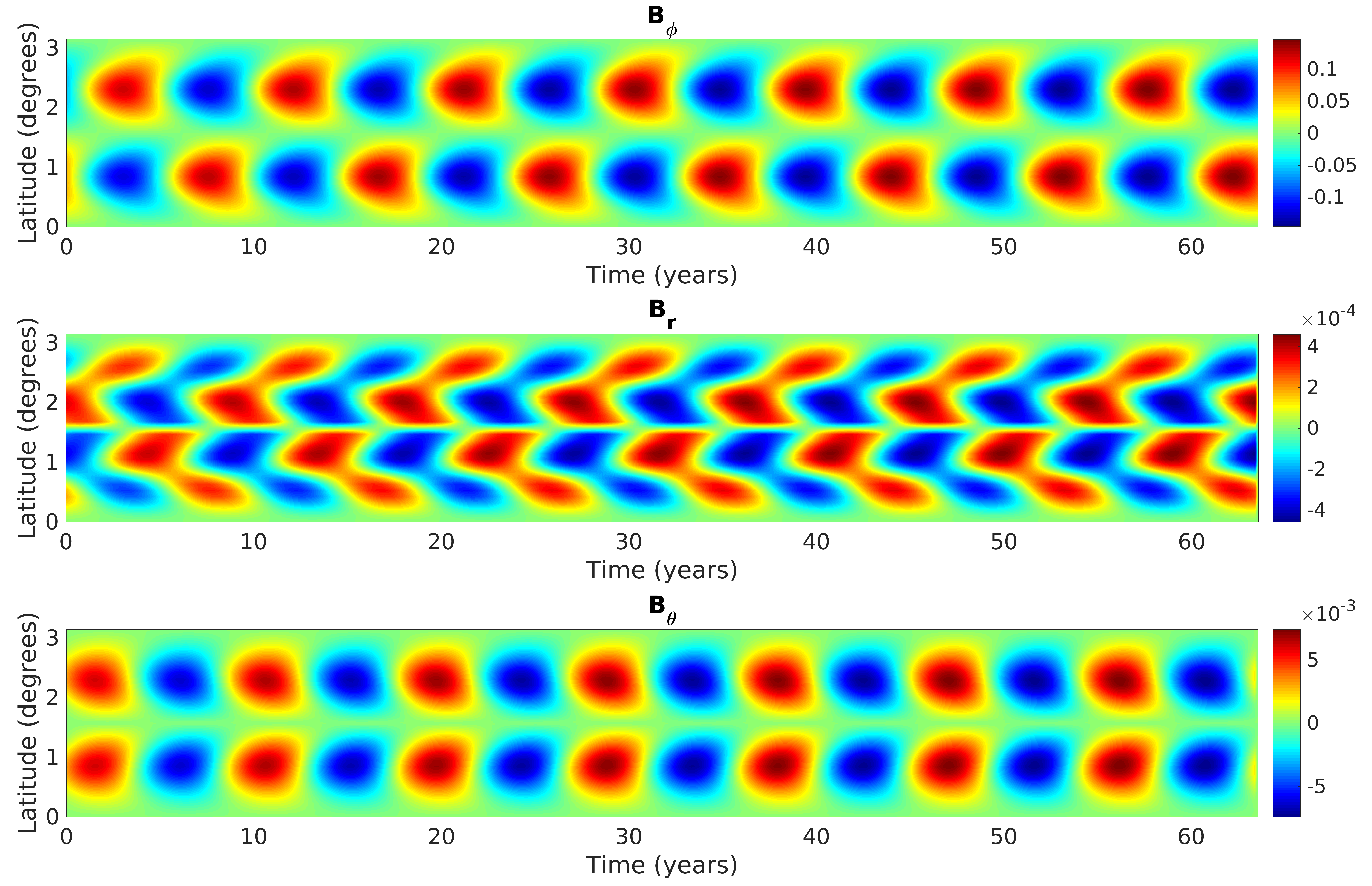}
      \captionof{subfigure}{Case $2\,\omega_\odot$}
      \label{fig:campo2w}
    \end{minipage}%
    }
  }
  \caption{Toroidal ($B_\phi$), radial ($B_r$), and polar ($B_{\theta}$) magnetic fields as a function of time for different rotation rates.}
  \label{B}
\end{figure}

To simulate these six solar twins, we kept all solar parameters fixed and varied only their rotation periods, obtaining different activity periods for each case (Figure~\ref{Metcalfe+hd}). Simulations were carried out for rotation periods between $0.5\omega_\odot$ and $3\omega_\odot$. Figure~\ref{B} shows the evolution of the three components of the magnetic field at the stellar tachocline derived from the described model for half and double solar rotation periods. From visual inspection, different cycle lengths are appreciable in both figures; the magnetic maps for  $0.5\omega_\odot$ exhibit an activity cycle of $\sim$ 20 years, while the maps corresponding to  the $2\omega_\odot$-rotation rate are related to a $\sim$ 5-year activity cycle.

\section{Periods of Activity vs the Rotation Period.}

Table~\ref{resumen} compiles the main characteristics of the stars in the studied sample that presented activity cycles, along with the results obtained in this work.

\begin{table}[htb!]
\centering
\caption{Activity cycles in solar-type stars. Selection of solar-type stars from the HK$\alpha$ Project (with 3 days $<P_{\mathrm{rot}}<$ 60 days), their main parameters, the years of observation at CASLEO, the number of CASLEO observations used, and the results obtained in this work. Rotation periods were taken exclusively from the literature (a--\citealt{Cincunegui07}, b--\citealt{Metcalfe10}, c--\citealt{Metcalfe13}, d--\citealt{Egeland15}).}
\resizebox{0.9\textwidth}{!}{%
\begin{tabular}{lccccc}
\hline
\textbf{ID} & \textbf{Spectral Type} & \textbf{Time Proc} & \textbf{P$_{\mathrm{rot}}$ (d)} & \textbf{P$^{1}_{act}$ (y)} & \textbf{P$^{2}_{act}$ (y)} \\
\hline
HD 59967  & G3 V & 2002--2023 & 10.26$^{a}$ & --   & 1.17 $\pm$ 0.03  \\
HD 38392  & K2 V & 2000--2022 & 17.3$^{a}$  & 17.1 $\pm$ 1.8 & 2.75 $\pm$ 0.09 \\
HD 3443   & K1 V & 2000--2022 & 30$^{a}$ & 2.90 $\pm$ 0.05 & --    \\
HD 32147  & K3 V & 2000--2023 & 47.4$^{a}$ & 10.7 $\pm$ 0.6 & 3.95 $\pm$ 0.09  \\
HD 30495  & G3 IV-V & 2000--2022 & 19.36$^{d}$ & 18.5 $\pm$ 3.6 & 4.7 $\pm$ 0.2  \\
HD 27442  & K2 III & 2001--2022 & 54.86$^{a}$ & 19.4 $\pm$ 3.2 & --    \\
HD 26965  & K0 V & 2000--2020 & 37.1$^{a}$  & 9.3 $\pm$ 0.3 & --    \\
HD 22049  & K2 V  & 2000--2022 & 11.11$^{c}$ & --   & 2.93 $\pm$ 0.05  \\
HD 219834 & G8 IV & 2000--2022 & 42.5$^{a}$ & 20.2 $\pm$ 1.5 & -- \\
HD 216803 & K4 V & 2003--2022 & 10.3$^{a}$ & 3.4 $\pm$ 0.1 & --    \\
HD 187923 & G0 V & 2000--2014 & 3.72$^{a}$ & 1.17 $\pm$ 0.02 & --    \\
HD 1835   & G3 V & 2000--2021 & 7.78$^{a}$  & --   & 1.47 $\pm$ 0.02  \\
HD 17925  & K1 V & 2000--2022 & 6.76$^{a}$  & 17.1 $\pm$ 2.3 & 0.690 $\pm$ 0.004  \\
HD 17051  & G0 V  & 2002--2022 & 8.5$^{b}$ & 1.4 $\pm$ 0.02 & --    \\
HD 16673  & F6 V  & 2000--2021 & 7.0$^{a}$ & --   & 0.93 $\pm$ 0.01  \\
HD 160691 & G3 IV-V & 2000--2022 & 14.04$^{a}$ & 22.1 $\pm$ 3.1 & 2.23 $\pm$ 0.03  \\
HD 152391 & G8 V  & 2000--2018 & 11.43$^{a}$ & 8.6 $\pm$ 0.6 & 19.8 $\pm$ 2.5  \\
HD 147513 & G5 V  & 2000--2022 & 8.5$^{a}$ & 8.11 $\pm$ 0.5 & 1.37 $\pm$ 0.01  \\
HD 13445  & G3 IV-V & 2002--2022 & 30$^{a}$ & 22.3 $\pm$ 3.9 & --    \\
\hline
\end{tabular}
}
\label{resumen}
\end{table}

\begin{figure}[hbt!]
\centering
\includegraphics[width=0.65\columnwidth]{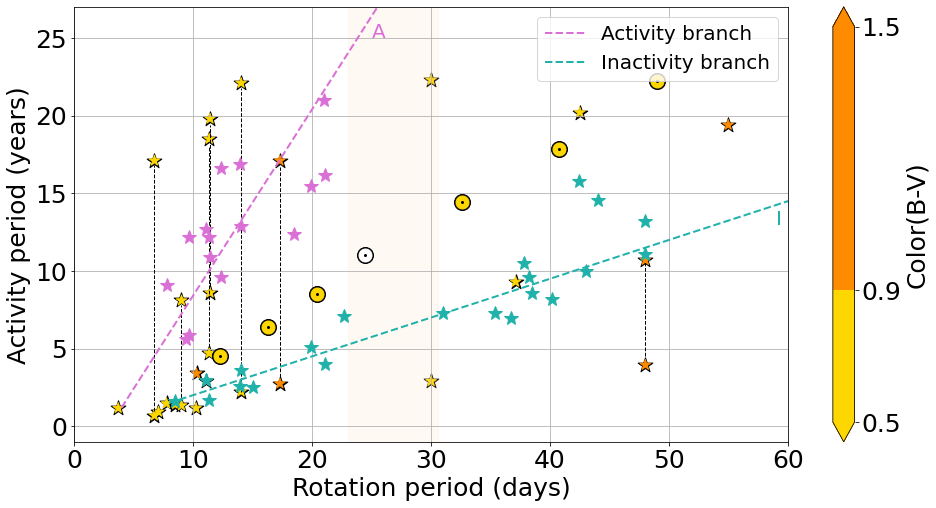}
\caption{Stars belonging to the active branch (A) are indicated in magenta and those of the inactive branch (I) in cyan \citep{Saar99,Metcalfe10,Metcalfe13,Hall07,BohmVitense07,Egeland15}. The dotted magenta and cyan lines are the linear fit reported by \cite{BohmVitense07}, while the Sun is indicated with its typical symbol. The symbols joined with dotted lines correspond to activity periods of the same star. The yellow and orange ones are the results obtained in the present work listed in Table \ref{resumen}. Based on the model developed in \cite{Sraibman19}, we built solar dynamos for different rotation periods. In this figure the simulated cycles are shown with the a yellow classical solar symbol.}
\label{Metcalfe+hd}
\end{figure}

Upon determining reliable rotation and activity periods for the target stars, we proceeded to construct a diagram (Figure \ref{Metcalfe+hd}) correlating the activity cycle length with the stellar rotation rate, incorporating data from previous studies \citep{Saar99,Metcalfe10,Metcalfe13,Hall07,BohmVitense07,Egeland15}.

Among the set of 44 stars analysed, activity cycles were detected in 19 stellar objects: HD 59967, HD 38392, HD 3443, HD 32147, HD 30495, HD 27442, HD 26965, HD 22049, HD 219834, HD 216803, HD 187923, HD 1835, HD 17925, HD 17051, HD 16673, HD 160691, HD 152391, HD 13345, HD 147513.  Six of these stars were previously studied by \cite{Metcalfe16} and are included in the original $P_{\mathrm{cyc}}$--$P_{\mathrm{rot}}$ diagram (magenta and cyan stars). The cycle of HD 17051, the short cycle of HD 22049 and the long cycle of HD 32147 seems to coincide with the reported cycle, while the long cycle of HD 22049 reported in \cite{Metcalfe13}  was not detected and the short cycle of HD 152391   is 10\% lower than the one obtained in \cite{Saar99}. However, a short cycle was found for HD 1835, not previously reported in the literature. The observed differences are presumably a consequence of the differing time span and sampling density of our dataset. Also a non-previously reported short cycle  for HD 3217  and long cycle for HD 152391 were detected. Similarly, we found a single long activity cycle for HD 30495, which is outside the range of periods reported by \cite{Egeland15}. A more exhaustive analysis will be necessary to reach a definitive conclusion.
Moreover, in Figure \ref{Metcalfe+hd} we can see that there are three stars that present long cycles that exceed the values expected from the active branch that could eventually belong to the superactive branch considered in \cite{Saar99}. Finally, simulated solar cycles seem to place between both branches for slow and fast rotator suns, consistent with the 11-yr solar cycle.

\bibliographystyle{unsrtnat}
\small
\bibliography{Sample}

@article{Metcalfe16,
       author = {Metcalfe, T. S. and Egeland, R. and van Saders, J.},
        title = "{Stellar Evidence That the Solar Dynamo May Be in Transition}",
      journal = {ApJL},
         year = 2016,
        month = jul,
       volume = {826},
       number = {1},
          doi = {10.3847/2041-8205/826/1/L2},
archivePrefix = {arXiv},
       eprint = {1606.01926},
 primaryClass = {astro-ph.SR},
       adsurl = {https://ui.adsabs.harvard.edu/abs/2016ApJ...826L...2M}
}

@article{Zechmeister09,
   author = {Zechmeister, M. and K{\"u}rster, M.},
    title = "{The generalised Lomb-Scargle periodogram. A new formalism for the floating-mean and Keplerian periodograms}",
  journal = {A{\&}A},
archivePrefix = "arXiv",
   eprint = {0901.2573},
 primaryClass = "astro-ph.IM",
     year = 2009,
    month = mar,
   volume = 496,
    pages = {577-584},
      doi = {10.1051/0004-6361:200811296},
   adsurl = {http://adsabs.harvard.edu/abs/2009A\%26A...496..577Z}
}

@article{Lovis11,
   author = {Lovis, C. and Dumusque, X. and Santos, N.~C. and Bouchy, F. and Mayor, M. and Pepe, F. and Queloz, D. and S{\'e}gransan, D. and Udry, S.},
    title = "{The HARPS search for southern extra-solar planets. XXXI. Magnetic activity cycles in solar-type stars: statistics and impact on precise radial velocities}",
  journal = {ArXiv e-prints},
archivePrefix = "arXiv",
   eprint = {1107.5325},
 primaryClass = "astro-ph.SR",
     year = 2011,
    month = jul,
   adsurl = {http://adsabs.harvard.edu/abs/2011arXiv1107.5325L}
}

@article{Jenkins08,
   author = {Jenkins, J.S. and Jones, H.R.A. and Pablenko, Y. and Pinfield, D.J. and Barnes, J.R. and Lyubchik, Y.},
    title = "{Metallicities \& Activities of Southern Stars}",
  journal = {A{\&}A},
archivePrefix = "arXiv",
     year = 2008,
    month = oct,
    volume = 458,
    pages = {571-584},
      doi = {10.1051/0004-6361:20078611},
   adsurl ={https://ui.adsabs.harvard.edu/abs/2008A%26A...485..571J/abstract}
}

@article{Cincunegui07,
   author = {Cincunegui, C. and D{\'{\i}}az, R.~F. and Mauas, P.~J.~D.},
    title = "{A possible activity cycle in Proxima Centauri}",
  journal = {A{\&}A},
   eprint = {astro-ph/0703514},
     year = 2007,
    month = jan,
   volume = 461,
      doi = {10.1051/0004-6361:20066027},
   adsurl = {http://adsabs.harvard.edu/abs/2007A%26A...461.1107C}
}

@article{Metcalfe10,
       author = {Metcalfe, T.~S. and Basu, S. and Henry, T.~J. and Soderblom, D.~R. and Judge, P.~G. and Kn{\"o}lker, M. and Mathur, S. and Rempel, M.},
        title = "{Discovery of a 1.6 Year Magnetic Activity Cycle in the Exoplanet Host Star {\ensuremath{\i}} Horologii}",
      journal = {ApJL},
         year = 2010,
        month = nov,
       volume = {723},
       number = {2},
        pages = {L213-L217},
          doi = {10.1088/2041-8205/723/2/L213},
archivePrefix = {arXiv},
       eprint = {1009.5399},
 primaryClass = {astro-ph.SR},
       adsurl = {https://ui.adsabs.harvard.edu/abs/2010ApJ...723L.213M}
}

@article{Ibanez19a,
   author = {Iba{\~n}ez Bustos, R.~V. and Buccino, A.~P. and Flores, M. and Martinez, C.~I. and Maizel, D. and Messina, S. and Mauas, P.~J.~D.},
    title = "{First long-term activity study of AU Microscopii: a possible chromospheric cycle}",
  journal = {MNRAS},
     year = 2019,
    month = feb,
   volume = 483,
    pages = {1159-1167},
      doi = {10.1093/mnras/sty3147},
   adsurl = {http://adsabs.harvard.edu/abs/2019MNRAS.483.1159I}
}

@ARTICLE{Ibanez19b,
       author = {{Iba{\~n}ez Bustos}, R.~V. and {Buccino}, A.~P. and {Flores}, M. and {Mauas}, P.~J.~D.},
        title = "{Ross 128 - GL 447. A possible activity cycle for a slow-rotating fully convective star}",
      journal = {A{\&}A},
         year = 2019,
        month = aug,
       volume = {628},
          eid = {L1},
        pages = {L1},
          doi = {10.1051/0004-6361/201936030},
archivePrefix = {arXiv},
       eprint = {1907.05728},
 primaryClass = {astro-ph.SR},
       adsurl = {https://ui.adsabs.harvard.edu/abs/2019A&A...628L...1I}
}

@article{Ibanez20,
   author = {Iba{\~n}ez Bustos, R.~V. and Buccino, A.~P. and Messina, S. and Lanza, A.F. and Mauas, P.~J.~D.},
    title = "{Activity–rotation in the dM4 star Gl 729. A possible chromospheric cycle}",
  journal = {A{\&}A},
     year = 2020,
    month = dec,
   volume = 644,
      doi = {10.1051/0004-6361/202039164},
}

@article{Metcalfe13,
   author = {Metcalfe, T.~S. and Buccino, A.~P. and Brown, B.~P. and Mathur, S. and Soderblom, D.~R. and Henry, T.~J. and Mauas, P.~J.~D. and Petrucci, R. and Hall, J.~C. and Basu, S.},
    title = "{Magnetic Activity Cycles in the Exoplanet Host Star epsilon Eridani}",
  journal = {ApJL},
archivePrefix = "arXiv",
   eprint = {1212.4425},
 primaryClass = "astro-ph.SR",
     year = 2013,
    month = feb,
   volume = 763,
      eid = {L26},
    pages = {L26},
      doi = {10.1088/2041-8205/763/2/L26},
   adsurl = {http://adsabs.harvard.edu/abs/2013ApJ...763L..26M}
}

@article{Sraibman19,
   author = {Sraibman, L. and Minotti, F.},
    title = "{Large-scale Model of the Axisymmetric Dynamo with Feedback Effects}",
  journal = {Solar Physics},
     year = 2019,
    month = jan,
   volume = 294,
      eid = {14},
    pages = {14},
      doi = {10.1007/s11207-018-1350-1},
   adsurl = {http://adsabs.harvard.edu/abs/2019SoPh..294...14S}
}

@article{BohmVitense07,
   author = {B{\"o}hm-Vitense, E.},
    title = "{Chromospheric Activity in G and K Main-Sequence Stars, and What It Tells Us about Stellar Dynamos}",
  journal = {ApJL},
     year = 2007,
    month = mar,
   volume = 657,
    pages = {486-493},
      doi = {10.1086/510482},
   adsurl = {http://adsabs.harvard.edu/abs/2007ApJ...657..486B}
}

@article{Egeland15,
   author = {Egeland, R. and Metcalfe, T.~S. and Hall, J.~C. and Henry, G.~W.},
    title = "{Sun-like Magnetic Cycles in the Rapidly-rotating Young Solar Analog HD 30495}",
  journal = {ApJL},
archivePrefix = "arXiv",
   eprint = {1507.03611},
 primaryClass = "astro-ph.SR",
     year = 2015,
    month = oct,
   volume = 812,
      eid = {12},
    pages = {12},
      doi = {10.1088/0004-637X/812/1/12},
   adsurl = {http://adsabs.harvard.edu/abs/2015ApJ...812...12E}
}

@article{Vaughan78,
   author = {Vaughan, A. and Preston, G. and Wilson, O.},
    title = "{Flux Measurements of CA II H and K emission}",
  journal = {PASP},
archivePrefix = "arXiv",
   eprint = {1312.3810},
 primaryClass = "astro-ph.SR",
     year = 1978,
   volume = 90,
    pages = {267-274},
      doi = {}
}

@ARTICLE{Baum22,
  author  = {Baum, A. C. and Wright, J. T. and Luhn, J. K. and Isaacson, H.},
  title   = {Five Decades of Chromospheric Activity in 59 Sun-like Stars and New Maunder Minimum Candidate HD 166620},
  journal = {AJ},
  year    = {2022},
  volume  = {163},
  pages   = {183}
}

@article{Saar99,
       author = {{Saar}, Steven H. and {Brandenburg}, Axel},
        title = "{Time Evolution of the Magnetic Activity Cycle Period. II. Results for an Expanded Stellar Sample}",
      journal = {ApJ},
         year = 1999,
        month = oct,
       volume = {524},
       number = {1},
        pages = {295-310},
          doi = {10.1086/307794},
       adsurl = {https://ui.adsabs.harvard.edu/abs/1999ApJ...524..295S}
}

@article{Hall07,
       author = {{Hall}, Jeffrey C. and {Lockwood}, G.~W. and {Skiff}, Brian A.},
        title = "{The Activity and Variability of the Sun and Sun-like Stars. I. Synoptic Ca II H and K Observations}",
      journal = {AJ},
         year = 2007,
        month = mar,
       volume = {133},
       number = {3},
        pages = {862-881},
          doi = {10.1086/510356},
       adsurl = {https://ui.adsabs.harvard.edu/abs/2007AJ....133..862H}
}

@article{Colombo24,
       author = {{Colombo}, P.~D. and {Buccino}, A.~P. and {Oviedo}, C.~G. and {Iba{\~n}ez Bustos}, R. and {Martinez}, C.~F. and {Mauas}, P.},
        title = "{Estudio preliminar de ciclos de actividad en estrellas del tipo solar}",
      journal = {BAAA},
         year = 2024,
        month = aug,
       volume = {65},
        pages = {71-74},
       adsurl = {https://ui.adsabs.harvard.edu/abs/2024BAAA...65...71C}
}

@ARTICLE{Minnoti00,
       author = {{Minotti}, Fernando O.},
        title = "{Self-consistent derivation of subgrid stresses for large-scale fluid equations}",
      journal = {Physical Review E},
         year = 2000,
        month = jan,
       volume = {61},
       number = {1},
        pages = {429-434},
          doi = {10.1103/PhysRevE.61.429},
       adsurl = {https://ui.adsabs.harvard.edu/abs/2000PhRvE..61..429M}
}

\end{document}